# The politics of deceptive borders: 'biomarkers of deceit' and the case of iBorderCtrl

Javier Sánchez-Monedero[a]* and Lina Dencik[a]

[a]School of Journalism, Media and Culture, Cardiff University, Cardiff, United Kingdom

*corresponding author: Javier Sánchez-Monedero. Cardiff University, School of Journalism, Media and Culture. Two Central Square, Central Square, CARDIFF CF10 1FS. Sanchez-monederoJ@cardiff.ac.uk

**Abstract**: This paper critically examines a recently developed proposal for a border control system called iBorderCtrl, designed to detect deception based on facial recognition technology and the measurement of micro-expressions, termed 'biomarkers of deceit'. Funded under the European Commission's Horizon 2020 programme, we situate our analysis in the wider political economy of 'emotional AI' and the history of deception detection technologies. We then move on to interrogate the design of iBorderCtrl using publicly available documents and assess the assumptions and scientific validation underpinning the project design. Finally, drawing on a Bayesian analysis we outline statistical fallacies in the foundational premise of mass screening and argue that it is very unlikely that the model that iBorderCtrl provides for deception detection would work in practice. By interrogating actual systems in this way, we argue that we can begin to question the very premise of the development of data-driven systems, and emotional AI and deception detection in particular, pushing back on the assumption that these systems are fulfilling the tasks they claim to be attending to and instead ask what function such projects carry out in the creation of subjects and management of populations. This function is not merely technical but, rather, we argue, distinctly political and forms part of a mode of governance increasingly shaping life opportunities and fundamental rights.

## Introduction

As data-centric technologies come to shape more and more of social life, the areas of borders and migration management have become prominent as sites of experimentation and investment. Characterised in part by a post-9/11 securitisation logic, the so-called refugee 'crisis' of 2015 has heightened focus on border politics, not least in Europe where the simultaneous externalisation and internalisation of borders continue to shape the geopolitics of the European project. The European Commission has set aside a proposed €34.9 billion for border control and migration management between 2021 and 2027 (Gallagher & Jona, 2019). The generation and collection of data plays a pertinent role in this context, with vast interoperable databases, digital registration processes, biometric data collection, social media identity verification, and various forms of data-driven risk and vulnerability assessments now a key part of European border regimes (Metcalfe & Dencik, 2019). Information systems such as VIS, SIS, and EURODAC operate to control the border crossing traffic, migration and asylum applications and electronic passports, facial recognition technologies and other biometric information make up the advent of so-called 'smart borders' that have become the hallmarks of EU-funded research and development projects in recent years (Cannataci, 2016).

These 'smart borders' not only serve new forms of identification and categorisation of people on the move. Increasingly, we are also seeing the (re)emergence of recognition technologies used for detecting emotion, affective states (such as stress) and deception as part of a growing risk assessment industry organised around a logic of data accumulation. In 2011, the UK Border Agency deployed an operational trial to evaluate stress, anxiety and deception at the immigration desk based on a facial and thermal analysis tool developed by researchers together with defence technology company QinetiQ and funded by the UK Engineering and Physical Sciences Research Council (EPSRC) (POST, 2011; Ugail et al., 2007). AVATAR, another automated lie detector based on eye-tracking, has been tested on the southern US border (F. Nunamaker et al., 2013) and in 2012, the same team that commercialized AVATAR performed field experiments with EU border guards from several countries to test an early version of the lie detector (Elkins et al., 2012).

Whilst little is known about the outcomes of these pilot programmes, investment in border control technologies said to identify deception and risk has continued. Recent attention has particularly focused on the controversial EU Horizon 2020-funded project iBorderCtrl (Intelligent Portable Control System), which describes its aims as deploying 'well established as well as novel technologies together to collect data that will move beyond biometrics and onto biomarkers of deceit' piloted at European borders (iBorderCtrl, 2016a). Dismissed as 'pseudo-science' (Boffey, 2018) and subject to considerable counter-research and activism (see for example the initiative *iBorderCtrl.no*), the iBorderCtrl project is a significant case study for engaging with the politics of data-driven technologies. In particular, it highlights the relevance of a booming industry in 'empathic media' and 'emotional AI' (McStay, 2018) as it becomes intertwined with highly securitised policy agendas pursued in a context of perceived crisis.

In this article, we build on research into deception detection and risk assessments in order to comprehensively illustrate the politics of both design and execution of the iBorderCtrl project. We start by situating the iBorderCtrl project in the context of data-driven governance and the growth of recognition technologies in the digital economy, before moving on to outlining the field of deception detection and its current incarnation in an age of machine learning (ML) as the most popular technique to build artificial intelligence (AI) systems. This forms the backdrop to our analysis of the iBorderCtrl project and the technopolitical assumptions that underpin it. We draw on publicly available documentation in order to provide a statistical evaluation of how the method used in the iBorderCtrl project would work in practice. In particular, we outline the central limitation of applying test design based on data points from a control group to the general population. This statistical limitation in mass screenings is a feature (rather than a bug) of the performance of recognition technologies, that has significant implications for the rights of people, especially marginalised groups and vulnerable populations that are often least able to challenge such systems. In arguing this, we make the case for understanding the pursuit of 'smart borders' and the nature of the technologies that make up the current European border regime as distinctly political processes that need to be engaged with as part of a particular mode of governance.

**The political economy of iBorderCtrl**

The advent of deception technologies has a long history, but has gained prominence in recent years in conjunction with the growth of 'empathic media', particularly in the form of 'emotional AI' that involves, according to McStay (2018), 'reading words and images, seeing and sensing facial expressions, gaze direction, gestures and voice. It also encompasses machines feeling our heart rate, body temperature, respiration and the electrical properties of our skin, among other bodily behaviours.' The term 'emotional AI' is closely related to the classic field of 'affective computing' and is increasingly used by industry. It is a way of engaging with emotion as something that can be observed through means of what can be surveyed, measured and remembered (rather than any 'mentalistic' process) that works well with sensing techniques that classify facial and bodily behaviour. In particular, these are sensing techniques that have gained prominence in a context of datafication that refers to the trend of turning increasing aspects of social phenomena and human behaviour into quantified formats that can be tabulated and analysed (Mayer-Schönberger & Cukier, 2013). Whilst much of this datafication initially focused on metadata based on communication and online activities, it increasingly extends to sensors and facial recognition software that generates data based on movements, expressions and physiology, significantly blurring the contours of public and private experiences, and amplifying the embodied construction of data subjects.

The use of such technologies is rapidly becoming a feature of commercial services, including digital platforms and devices associated with the Internet of Things, as well as in the workplace (Sánchez-Monedero & Dencik, 2019). States and governments, often in partnership with industry, have propelled the trend further, turning to data-driven recognition and detection technologies to enact governance. In Europe, the deployment of facial recognition technologies by police has sparked significant debate on citizen rights in public spaces, furthered by a growing discourse on "smart cities" (see for example the Face Off campaign from Big Brother Watch). Education is another area where facial recognition tools are being experimented with (McStay, 2019), drawing on developments in China where these are used, amongst other things, to assess attention levels amongst school pupils (Connor, 2018). Yet it is perhaps no surprise that it is particularly in border control, and migration management more broadly, where European states have sought to increase investment in technological tools at significant speed.

The socio-technical assemblage that now makes up border regimes, what Pötzsch (2015) aptly refers to as the advent of the "iBorder" as a way to articulate the dispersal of the border into remote, algorithmic decisions capable of determining risks, serves as an important setting for the development of iBorderCtrl. Proposed by researchers at Manchester Metropolitan University in the UK, the iBorderCtrl project has been supported by a €4.5 million research grant under the Horizon2020 programme at the European Research Council. It continues a long-standing history of EU-funded security technology, not least in the context of border control and migration management, that has gained further prominence with the growing enthusiasm for the potentials of ML and AI. Significantly, there have been concerns about a lack of transparency surrounding the processes and details of the iBorderCtrl project, including ethics questions and the relationship between the research team and private companies on the project (Wilde 2018; Gallagher and Jona 2019).

The iBorderCtrl project centres on the ability to perform automatic "deception detection" and "risk assessment" in the border-crossing encounter. The project proposes a two-stage process with a pre-registration step to provide traveller information, and a later border crossing stage that includes biometric identification and matching, document authenticity analysis, interaction with external legacy and social systems, an Automatic Deception Detection System (ADDS), a Risk Based Assessment Tool (RBAT) and a post hoc analytics tool (iBorderCtrl, 2016b). The aim of the project is to speed up border control for third-country nationals crossing land borders of EU Member states by providing a decision support system for border authorities (iBorderCtrl, 2016c). The project includes pilot tests in Hungarian, Greek and Latvian land borders.

The features of iBorderCtrl are typical in smart border systems, but the project is novel in its application of the ADDS. In seeking to 'move beyond biometrics and onto biomarkers of deceit.', the project claims to build proxy features that represent, according to the authors, the act of 'deceiving' that will inform the categorization of persons as 'bona fide' and 'non-bona fide' travellers (iBorderCtrl, 2016b). As we will go on to detail further below, in epitomising not only the ideology of 'dataism' (Van Dijck, 2014) in its assumptions about the relationship between people and data, but also the simultaneous scientific void of affective recognition systems (Wilde, 2018) and 'erasure of doubt' in machine learning algorithms (Amoore, 2019), iBorderCtrl emerges as a paragon for the politics of data-driven governance. As a way to illustrate this, we start by situating our analysis of iBorderCtrl in a broader discussion of lie detection devices.

**Deception detection**

*Lie detectors*

Deception or lie detection devices, mainly represented by the polygraph, have always been contentious and lacking in substantial scientific evidence (Damphousse et al., 2007; National Research Council, 2003; Wilde, 2014). Generally, proposals on lie detectors assume that, during an interview, deceptive answers will produce unique cues such as physiological responses, facial expressions, pitch changes amongst others (DePaulo et al., 2003) that can be accurately measured. A second important assumption shared by many studies and proposals is that lying is a valid predictor of guilt (Rajoub & Zwiggelaar, 2014).

Although scientific consensus is that the accuracy of the polygraph and other devices is close to chance when rigorous experimental criteria are met (Saxe & Ben-Shakhar, 1999), industry proponents of the polygraph claim that deceptive answers produce physiological responses that can be detected by monitoring blood pressure, pulse, respiration, and/or skin conductivity (Wilde 2014). A review from the US National Research Council (NRC) (2003) concluded that the polygraph, even if it worked as its vendors claim, would not be useful to perform population screenings with low rates of events in the population since it would produce a large number of false positives (we elaborate on this below). In addition, the study warns that the indicators these tests rely on can be gamed through cognitive or physical means. What is more, in a meta-analysis of 158 cues associated with deception, including nonverbal cues, DePaulo et al. (2003), concluded that the majority of cues are unreliable. Nevertheless, the use of lie detectors is relatively widespread in some US sectors such as insurance, although in most parts of the world it is not recognized as a valid test (cf. Wilde, 2014).

Beyond the polygraph, there have been other devices based on Functional Magnetic Resonance Imaging (fMRI) (Simpson, 2008), eye-tracking (F. Nunamaker et al., 2013), electroencephalography (EEG) (Heussen et al., 2010), voice analysis (NITV Federal Services, 2019) or thermal facial analysis (Rajoub & Zwiggelaar, 2014) among others. Many of these methods have been scrutinized by third parties who tend to disagree with the reported accuracy from developers of the devices and conclude that the experimental validations were generally weak. Often the number of participants for testing the tool is far too low to make any robust claims about its accuracy (Heussen et al., 2010). With regards to voice analysis, the NRC in the US concluded that the studies validating voice stress analysis (VSA) as a means of lie detection 'offer little or no scientific basis for the use of the computer voice stress analyser or similar voice measurement instruments' (NRC, 2003). Later work by linguistic scholars Eriksson and Lacerda (2008) studied the feasibility of computer-based VSA (CVSA) and concluded that there was no scientific evidence to support the manufacturers' claims. Rigorous experimental studies found that the detection accuracy was worse than chance level: CVSA correctly identified only 15% of deceptive subjects on average (Damphousse et al., 2007). The product analysed by Eriksson and Lacerda (2008) was tested in the UK Department for Work and Pensions over three years. In 2011 the department concluded the detector was not useful after spending £2.4 million (Lomas, 2010).

*Data-driven deception detection*

Deception detection started a new era with the popularization of ML for assessment of psychometrics (Burr & Cristianini, 2019). Rather than performing an analysis of signals of the polygraph by trained experts, systems in this vein tend to rely on different types of data to fit statistical models to perform the classification task. The variety of systems typically extract features from raw data, such as video or other types of sensors, that represents a time window of the moment a person answers a question, and then label these vectors of features as deceptive or truthful. ML methods fit a model to the training dataset to map the multidimensional feature space to the output label space (deception or no-deception labels). The features can be engineered based on previous research, e.g. a feature that measures the blinking of each eye, or they can be learned by the model as part of the overall fitting process, as deep learning and autoencoders do.

ML is the ideal tool to create such mappings since many models such as artificial neural networks are universal approximators, meaning they can learn/build/fit any function to relate the input space with the label space provided the model is complex enough (Cybenko, 1989). Recent work has demonstrated that (deep) neural networks are also able to learn random labelling of data by memorizing the whole training set (C. Zhang et al., 2016). As we will elaborate on in our analysis of iBorderCtrl, since ML can build models that correlate any-input to any-output, data science practitioners have to be particularly careful to avoid issues such as overfitting, the representativeness of the sample, the reductionism of the optimization process or the misalignment of the ML optimization task with the domain optimization task (Lipton, 2018).

Considering the capacity of ML methods to fit models in complex datasets, deception detection relies heavily on the quality of such data sets. The quality of datasets in applied ML refers to how close the test environment is with the deployment scenario, but is also related to rules of robustness in statistical learning such as the number of patterns to comply with the Central

Limit Theorem, the number of patterns in relation to the number of features (see the "curse of dimensionality" discussion below), the stratification of the sample, the assumption about independent and identically distributed random variables (i.i.d.), etc. All these factors determine the capacity of the system to generalize with unseen data and the conditions under which this can be expected and ensured.

The case of deception detection has many of the attributes of a weak-defined task: people cannot universally define either the input or the labels in terms of data, the binarization of the problem is a hard reduction from the spectrum of emotions of a person to a binary label (i.e., that lying consists of a set of simulated emotions in a laboratory), and it would be complex to formulate a rule or set of rules to perform the input-output mapping. In addition, the modelling of the dependent variable is a case of inference of a latent variable, meaning it is not directly observable. Since ML allow for fitting any training data, but not to generalise from it unless conditions are met, it is particularly risky to draw conclusions about the performance of ML models with non-rigorous datasets and without strategies to mitigate overfitting (e.g. as done in current deep learning methods but is not the case with ADDS).

The ADDS component of iBorderCtrl is based on what the authors refer to as 'biomarkers of deceit' (iBorderCtrl, 2016b). These are non-verbal facial micro-expressions that, according to the authors, all humans produce when lying during an answer. Those signals when combined are therefore said to be able to act as predictors of lying. The biomarkers of deceit are encoded as 38 features, or channels, such as 'left eye blink', 'increase in face redness' or head movement directions (O'Shea et al., 2018). Each feature is extracted from a video segment of one second in which a person is lying or not when answering a question (although it is not clear from the available documentation how they generate features for the segment). The video is captured at 30 frames per second (fps) with a video resolution of 640x480. Each training pattern of the dataset consists of the vector of 38 features and the label that indicates truth or deception. The gender of the person is also encoded as one of the 38 features. To create the dataset, 32 participants (actors) were assigned a "truthful" or "deceptive" role to perform during the interview. Each participant had to answer 13 questions (see Table 1) with each answer segmented in many vectors. According to the authors, this procedure generated 86,586 vectors. The dataset consisted of 10 participants classified as having Asian/Arabic ethnic background and 22 as White European background, and 22 classified as male and 10 as female.

The interviews took place in a lab setting. The participant selection and role assignment criteria are not detailed in the public documentation. The "deceptive" actors produced both directed "truthful" and directed "false" answers. To create a directed "false" answer, all "deceptive" participants were given fake identities and a short life history and were instructed to pack a suitcase with standard items and then to introduce a prohibited item, such as a 'simulated drug package' that consists of 'soap powder in clear packet'.

Based on the dataset of participants, an artificial neural network (ANN) was fitted to the training data and the performance evaluated with patterns of people that were not used to train the model. Each question $q$ was assigned a score $D_q$ that consists of the averaged sum of the prediction $d_n$ of the ANN for each video segment of the question. The model has two parameters to filter segment scores $d_n$ in order to delete unknown facial expressions from calculation of $D_q$. In the task of detecting a deceptive scenario, the authors reported an accuracy rate of 73.66% when detecting deception and 75.55% when detecting truthfulness.

## Assumptions, dataset and experimental validation

Having outlined some of the background of deception detection and details of the iBorderCtrl model, we can begin to identify some key assumptions that extend a historically flawed area of research but one that despite errors and lack of evidence has persisted as a technique for assessing people. The desire to find a "scientific" method to detect lying, affective states and emotions is what drives the investment in deception detection and emotional AI more broadly. This desire or political will far outpaces the validity of the technology. Moreover, as we go on to discuss in this section, it ignores the contested assumptions upon which such technology is based that, as Amoore (2019) puts it, 'generates the parameters against which uncertainty will be adjudicated' such that a multiplicity of doubts in the model can be condensed to a single output beyond doubt (deception or truth). We base our analysis of such technopolitical assumptions on a number of academic papers relating to the design of iBorderCtrl as well as public documents available from the iBorderCtrl project website.

The central premise of the ADDS component in the iBorderCtrtl project is that some non-EU travellers will lie about their identity, financial means, destination, or other aspects of their travel plans that concern the state of desired entry at the point of entry. The project does not specify the intentions of travellers when providing fake information, but the example questions are focused on identity confirmation, the content of luggage and information related to other travellers, relatives or friends (see Table 1). According to the project's general scheme, a person identified as having likely given a deceptive response will be interviewed by a human border guard to perform further investigations.

*Table 1 Experiment interview questions from the ADDS validation from O'Shea et al. (2018).*

| Question Number | Question |
|---|---|
| 1 | What is your family name? |
| 2 | What is in your case? |
| 3 | Have you seen any posters of prohibited items? |
| 4 | Are there any items from the lists of prohibited items in your case? |
| 5 | How many items are in the case? |
| 6 | If you open the case and show me what is inside, will it confirm that your answers were true? |
| 7 | What is your first name? |
| 8 | When were you born? |
| 9 | Where were you born? |
| 10 | What is your current citizenship? |
| 11 | Please tell me the name of a friend or family member who can confirm your identity? |
| 12 | What is the relationship of this person to you? |
| 13 | Where does this person live? |

The goal of the iBorderCtrl smart border is to detect "bona fide" and "non-bona fide" travellers to identify illegal border crossings. Since the target is to discover illegal activities through lie detection, deception is considered a valid predictor of guilt and risk. Deception is indirectly observed though affective proxies such as anxiety that are indirectly observed by measuring non-verbal micro-expressions. Therefore, there is a chain of assumptions with regards to the relationship between criminality, lying, affective states and facial micro-expressions in which the role of ML is to fit a prediction model that can differentiate between deceptive and truthful responses. Based on information from public documents, it does not appear that this chain of assumptions is backed up with previous theoretical and empirical studies. The relation between deceptive behaviour and guilt and risk can be found in other pilot programs at borders: For example, AVATAR was tested by assigning 'guilty or innocent condition' to the participants and also notes that 'procedures were intended to heighten anxiety and simulate the circumstances surrounding actual criminal conduct' (Derrick et al., 2010). In the context of lie detection, the link between lie and guilt is generally direct, despite identified concerns that such a premise invites a sense of automatic suspicion amongst those subjected to the assessment (POST, 2011).

Another base assumption is that there exist non-verbal facial (micro) gestures, labelled as "biomarkers of deceit", that are indicators of deception during an interview. This assumption is based on the theories of Paul Ekman who claims that lying is an emotionally demanding task that may leave non-verbal behavioural traces (Ekman & Rosenberg, 2005). However, as noted above, reviews of studies suggest that assessment based on non-verbal behavioural observation is close to chance. Among several reasons, one simple explanation is that both people lying or telling the truth would do similar emotional work to appear honest when confronting an interview with consequences (Bond & DePaulo, 2006; Vrij & Granhag, 2012).

The model is also premised on the assumption that facial non-verbal micro-gestures can be measured. Micro-gestures or micro-expressions are extremely quick facial expressions that last between 1/25s and 1/5s and have been said to be used as proxy features to recognise emotions (M. Zhang et al., 2014). Micro-expressions are encoded into numerical features, for instance, by an algorithm that extracts a feature to measure eye blinking by tracking the eyes during a video segment. To capture such fast events, typically 200fps (frames per second) cameras are used (Davison et al., 2018; Yan et al., 2014). However, in a previous project by the iBorderCtrl team, Silent Talker, 15 fps cameras were used to capture micro-gestures (Rothwell et al., 2006) whereas the ADDS uses 30fps cameras (O'Shea et al., 2018). It is not clear from available documentation what the limitations of using such low fps cameras are for capturing micro-expressions.

The term "biomarker" or "biological marker" is typically used in medicine to refer to 'a broad subcategory of medical signs – that is, objective indications of medical state observed from outside the patient – which can be measured accurately and reproducibly' (Strimbu & Tavel, 2010). The team behind iBorderCtrl state that biomarkers of deceit are non-verbal signals that alone cannot reveal deceptive behaviour but together can be used by an ML method to detect lying. This means that according to this model, deceptive and non-deceptive behaviours are two non-overlapping categories that represent a set of emotional states. These states would be represented by the combination of 38 features (statistically correlated with the target). When assigning a label to the video segment(s) related to an answer, the ADDS only considers that the emotional states of a person can be grouped as deceptive or truthful and no additional possibilities are allowed, although the model has two parameters to filter un-meaningful segments when there is not a clear category for them.

In general, the common basis of lie detectors is that there are universal and involuntary physiological and physical responses that a person produces as a result of lying. In the case of the ADDS, it assumes that across persons, ethnicity, gender, age, functional diversity, neurodiversity, etc., there are universal ways of expressing deception through non-verbal expressions. According to the team, provided with a large and diverse enough training dataset, the tool might be able to detect deception amongst heterogeneous travellers (O'Shea et al., 2018). This assumption is partially shared with methods for emotion detection, which assume the existence of a universal commonality of expressions (McStay, 2018). What is more, the ADDS is premised on the assumption that the scenarios used for data collection and method validation are representative enough for real environment conditions, for example in the case of a 'simulated drug package' as noted above. Generating a suitable dataset is a common problem for data-driven deception detectors in which the experimental set-up consists of people interpreting a role. Previous projects have attempted to minimize artificiality and strengthen generalizability 'by offering substantial monetary bonuses' to participants judged by 'professional interviewers' as in the case of AVATAR, and by directing the participants to actually commit the crime in question (Derrick et al., 2010). Other attempts have included asking participants to interpret roles to convince an examiner that they were honest. As an exception, Pérez-Rosas et al. (2015) used real situations instead of actors by gathering data from real trial recordings, using the verdict of the trial and the testimonies verified by police to label clips as deceptive or truthful. Yet, apart from problems with representativeness stemming from the creation of situations of 'real' deception, the problems with relying on particular forms of behaviour as indicative of lying as well as the assumed universality of modes of deception cues across different demographics and populations remain significant fallacies in the training and validation scenarios.

Moreover, critical questions arise in relation to the way researchers attempted to validate the ADDS before conducting the pilot programs. The validation experiment provided by the team consisted of testing the performance of the ML classifier that would be included in the ADDS. This suggests that a test with new (unseen) individuals whose data needs to be captured, processed and classified by the ML model was not carried out. In this sense, there are questions about the actual extent of the test of the ADDS module.

As outlined above, the experimental setup for the ADDS component of the iBorderCtrl project consisted of a sample of 32 participants with each person generating 38-dimensional vectors for each question so that the training set consists of 86,586 training patterns. The generalization performance was tested with a leave-one-out (LOO) experimental design that was repeated 9 times to test the performance of the classifier in detecting deception in two unseen persons (one deceptive and one truthful participant). The detection rate for these 9 pairs of persons was, on average, 75.56% for deceptive and 73,67% for truthful participants. The standard deviation of the accuracy in the experiments with unseen persons was 24,37% for deceptive and 34,29% for truthful participants. The team explored the performance of the ADDS for persons whose vectors are both in the training and test sets. In this case, the mean performance is near 95% for both types of participants and the standard deviation is significantly reduced to 1% and 0.86% (see Table 2). Note that this second experiment cannot be justified from an experimental validation point of view.

The validation results present high variability in the classification performance of unseen participants. Since we do not have access to the dataset, we can only argue that the produced models may be overfitting because of the inherent problems of the training dataset, causing this instability in the predictions. The small sample size has some issues beyond the obvious lack of representation noted above, particularly in light of the nature of populations crossing European borders. From a statistical and ML point of view, the data sample needs to grow in relation to the number of input features (38), the number of classes (2) and the number of parameters in the model (unknown in this case). It can be argued that the number of data samples in the case of iBorderCtrl is 86,586; however, since these points are obtained from 30 participants (one truthful and one deceptive participant are reserved for testing), it is unclear whether the points extracted from each person can be rich enough to represent the population in the 38-dimensional feature space. The training dataset does not comply with the i.i.d. assumptions of most machine learning algorithms since the samples obtained from each participant are correlated and, thus, the generated patterns can be very close in the feature space and at the same time very distant to the vectors generated for other people. That is, the data in the feature space is very sparse, which can produce several problems. In statistics, this is known as the "curse of dimensionality" (COD) (Altman & Krzywinski, 2018) and it is especially relevant when the sample size is smaller than the number of dimensions of the data (30<38 under our hypothesis). The COD is well known to cause problems such as overfitting and unstable model predictions. Even without public information to fully check our hypothesis, we can see some hints to support it. As a consequence, we can conclude that it is very likely that the ML model overfits in relation to the data points and therefore achieves a better than chance performance in the validation experiments.

*Table 2 Performance of the ADDS in the validation experiment. Source Table IV in p.7 in O'Shea et al. (2018).*

|  | ADDS performance for unseen persons | | ADDS performance including the same person in the train and test datasets | |
| --- | --- | --- | --- | --- |
| Fold | T | D | T | D |
| 1 | 100.00 | 57.00 | 94.30 | 93.69 |
| 2 | 50.00 | 36.00 | 93.62 | 94.96 |
| 3 | 50.00 | 100.00 | 94.31 | 95.15 |
| 4 | 90.00 | 100.00 | 94.23 | 95.46 |
| 5 | 100.00 | 10.00 | 95.50 | 95.41 |
| 6 | 72.00 | 100.00 | 95.45 | 95.91 |
| 7 | 100.00 | 100.00 | 95.75 | 96.22 |
| 8 | 38.00 | 100.00 | 95.91 | 96.28 |
| 9 | 80.00 | 60.00 | 96.67 | 96.45 |
| 10 | - | - | 96.55 | 96.78 |
| Mean Acc. | 75.56 | 73.67 | 95.23 | 95.50 |
| STD Acc. | 24.37 | 34.29 | 1.00 | 0.86 |

These assumptions point to the contested scientific premises upon which iBorderCtrl has been able to position its model as a solution to a constructed problem of deception detection as a significant function of border control. These do not address the wider questions of the premises of optimisation in terms of risk estimation and speed of border crossings that are key features of the iBorderCtrl project (although not all detailed in the available publications); nor do they address the fundamental questions of the purpose of borders, the right to privacy, or the tendency towards discrimination against the most marginalised and vulnerable populations likely to experience the violence of borders.[1] Discussions on assumptions outlined above need to be considered in the context of these broader questions, but here we focus on some of the technical issues as a way to illustrate the politics of constructing problems and selling solutions in the context of a perceived security crisis.

---

[1] For a discussion of these issues in the context of iBortderCtrl, we refer to the resources published on www.iborderctrl.no

## Statistical limits of mass screening

In order to further elaborate on the limitations of the design of iBorderCtrl, we now turn our attention to assessing how the model presented in documentation would actually "work" in practice. Whilst our focus is specifically on the method for deception detection outlined in the iBorderCtrl project, we see our analysis as being relevant for models intended for mass screening more generally.

When performing a test to search for rare events, such as a deception detection classifier, in whole statistical populations, rather than directly considering the test outcome (i.e., the conditional probability of having a liar given the data), it is more suitable to draw conclusions from the expected predictive value, also known as posterior probability in Bayesian terminology. This type of analysis is common in medicine (Fenton & Neil, 2010) and public policy decisions and it is aligned with previous work evaluating mass screening proposals related to security (e.g. NRC 2003, Munk 2017). That is, this is an established way to calculate the probability that a person is an actual liar given the test output and the frequency of the event in the population, which is the relevant information for the decision-maker (the border guard in this case).

Bayesian statistics provide a principled method to incorporate prior knowledge in a domain, also named *beliefs*, about an event to evaluate the relationship between the observed data (test) and that belief. Another point of view is to separate the properties of a test or classifier from the characteristics of the statistical population. In this section, we will use the Bayes rule to interrogate the behaviour of the ADDS.

A test can produce four outcomes, in our case:
- True positive: the test detected deception and the person lied.
- False positive: the test detected deception and actually the person did not lie.
- True negative: the test does not detect deception and the person said the truth.
- False negative: the test does not detect deception and the person lied.

The performance of the test can be represented in the confusion matrix of conditional probabilities. Table 3 is built with the information provided in the documentation (O'Shea et al., 2018).

*Table 3 Confusion matrix of conditional probabilities of the ADDS. Source Table IV in p.7 in O'Shea et al. (2018).*

|  |  | Ground Truth | |
|---|---|---|---|
|  |  | *Lie* | *No-lie* |
| **Test Result** | *Test positive* | 73.66% | 24.45% |
|  | *Test negative* | 26.34% | 75.55% |

The test output produces conditional probabilities given the observed data (biomarkers of deception). For the above table, we interpret that if a person lies there is a 73.66% chance of having a positive test (true positive rate or sensitivity) and a 26.34% chance of having a negative test (false negative rate). If the person tells the truth, there is a 24.45% chance that the test will be positive (false positive rate) and a 75.55% chance that the test will be negative (true negative rate or specificity).

The framework of Bayesian statistics allows us to better understand the expected performance by incorporating knowledge about the context of the problem in the form of prior probability. The prior probability, also known as prevalence, represents the frequency of the event of interest (liars in our case) and regulates the level of trust in the prediction of a classifier in particular context. We can formulate a different hypothesis of the prior P(Lie), for instance, 5% of liars, and observe how each type of result of the test varies. For the case of 5% of liars, the corrected joint probabilities of each test outcome are shown in Table 4.

*Table 4 Joint probability matrix of expected performance of the ADDS for a context of 5% of liars (prior probability of 0.05).*

|  |  | **Ground Truth** | |
|---|---|---|---|
|  |  | *Lie (0.05)* | *No-lie (0.95)* |
| **Test Result** | *Test positive* | (True Pos.) 0.0368 | (False Pos.) 0.2323 |
|  | *Test negative* | (False Neg.) 0.0132 | (True Neg) 0.7177 |

The corrected performance is interpreted in the following way. For example, if we have 1,000 people being interviewed with 50 liars and 950 non-liars, ~38 out of 50 liars will be detected at the cost of wrongly labelling ~232 innocent people as liars. Approximately, 13 liars would not be detected by the test and ~717 people would be correctly classified as not deceptive.

Finally, to evaluate the expected behaviour of the deception test, Bayesian statistics help to answer the question *provided we have a positive test, what is the probability the person is lying?*, that formally corresponds to the posterior probability, also known as positive predictive value (PPV). We can calculate the posterior by using the Bayes Theorem:

$$P(\text{Lie} \mid +) = \frac{P(+ \mid \text{Lie})P(\text{Lie})}{P(+)} =$$

$$= \frac{P(+ \mid \text{Lie})P(\text{Lie})}{P(+ \mid \text{Lie})P(\text{Lie}) + P(+ \mid \text{Non-lie})P(\text{Non-lie})}$$

where:
- $P(\text{Lie} \mid +)$ is the probability of having a liar (Lie) given a positive test (+), this is the *posterior probability*.
- $P(+ \mid \text{Lie})$ is the chance of having a positive test (+) when the person is lying (Lie). This is the probability of true positive (73.66%).
- P(Lie) is the probability of having a person that would lie in the interview, formally *prior probability* but often referred to as *frequency* or *prevalence*. It is a characteristic of the population, not of the deception test.
- $P(\text{Non} - \text{lie})$ is the probability of having a person that is telling the truth (P(Non-lie)=1-P(Lie)).
- $P(+ \mid \text{Non-lie})$ is the probability of having a positive test (+) when the person is not lying (Non − lie). This is a false positive (24.45%).

Therefore, we can calculate the posterior with the hypothesis of 5% of liars in the population that will be screened:

$$P(\text{Liar} \mid +) = \frac{0.7366 \times 0.05}{0.7366 \times 0.05 + 0.2445 \times 0.95} \approx 13.69\%$$

This means that, in a scenario of 5% of liars, 13.69% of positive tests will correspond to actual liars crossing the borders and 86.31% will correspond to false positives, meaning only 1 positive in 8 persons labelled as a liar would correspond to an actual liar. Regarding false discoveries, the negative predictive value (NPV) or posterior of negative tests, $P(\text{Non-lie} \mid -)$, gives the probability of having actual truth-tellers when the deception test is negative. In the same scenario, the NPV is 98.20%, so most of the interviews with negative results correspond to people that did not lie. Figure 1 shows a graphic example of the decision tree of events relating to the corrected outcomes of the test and the posterior probabilities.

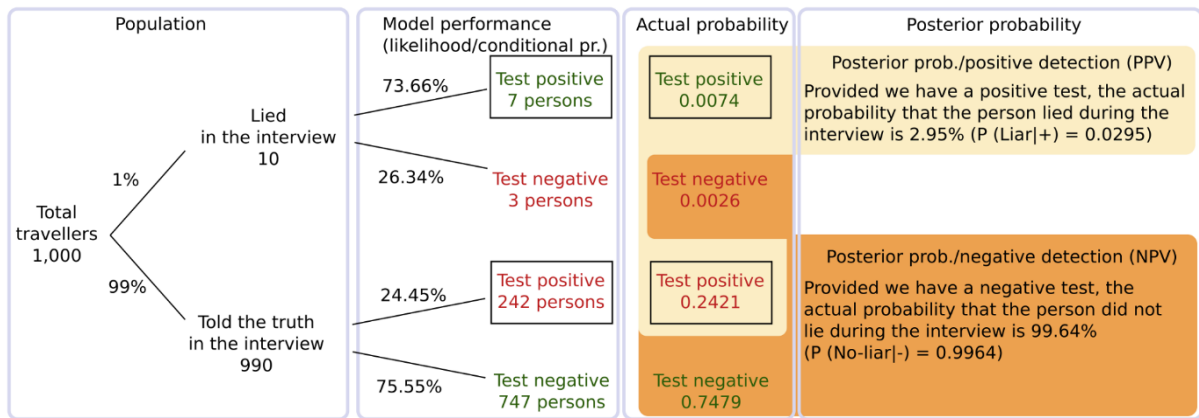

*Figure 1 Event tree to explain deception detection probability of each outcome. In this case we test the hypothesis of 1% of liars in the population*

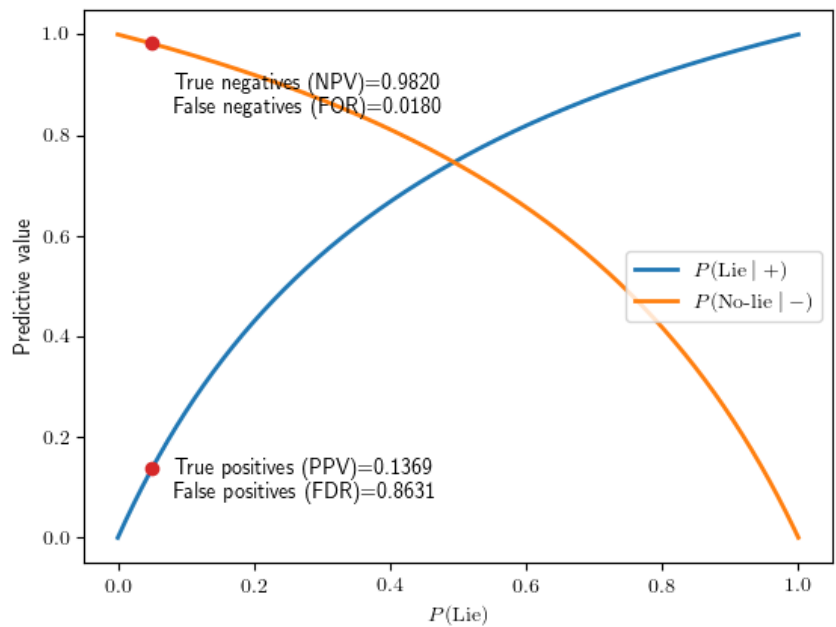

*Figure 2 Posterior probabilities of a positive and negative test for different scenarios of the frequency of deception. Highlighted values correspond to the scenario of 5% of liars. While the conditional probability is fixed, the probability of someone being a liar after a positive test depends on the general population.*

To highlight how the deception detection test would behave with different ratios of liars (prior probability), Figure 2 shows how the PPV and NPV vary under different hypotheses. Although it is not clear in the public documentation about iBorderCtrl how frequent it is believed that deception is during border-crossings, from Figure 2, we can know that if there is one liar or criminal in 10,000 people, the PPV would be 0.03%, if there is one in 1,000 the PPV would be 0.3% and 13.69% for the case of 5 in 100 people. It is straightforward to check that the smallest prior that produces a posterior of 50% would be the case of 1 criminal in every 4 persons.

As such, what this discussion highlights is that when transferred to a real-world context, even in the case that all the assumptions we identified in the previous section hold, the ADDS test will fail to identify many of the imposters while it simultaneously will flag many "bona fide" travellers as "liars". That is, even in the most favourable case for iBorderCtrl it is very unlikely that the tool can work in practice. This raises fundamental questions about the societal implications of such a system and who will be most impacted.

**Conclusion**

The turn to data-driven systems has taken hold of large swathes of state activity, often advanced by its own inherent logic of data accumulation as a recourse to better and more efficient forms of governance. Funding agendas and resource allocation is mimicking this logic, advancing projects that have sought to capitalise on the political economy of digital infrastructures and the increased securitisation of policy. The iBorderCtrl project epitomises the current moment of data politics in terms of its constitution, design and execution. Initiated in 2016 under the European Commission's Horizon 2020 programme, the project exemplifies the race to AI, the growing industry around biometrics and emotion detection for the purposes of population management, underpinned by a perceived political crisis that has strengthened the rhetoric of border regimes. By interrogating the actual mechanisms by which the iBorderCtrl project is said to function, we therefore want to draw attention to the wider politics of the development and deployment of such technologies. This is particularly pertinent as the "super-charged bureaucracy" (McQuillan, 2019) advanced by AI is part of (re)shaping the conditions of social, economic, and political injustice that are overwhelmingly burdened by the already marginalised (Benjamin, 2019; Browne, 2015). As McQuillan puts it, the inherent right to move and live is not a right if the person has to win it by confronting AI.

We have shown this by outlining the techno-politics of the iBorderCtrl project and the extension of deeply contested assumptions about the relationship between deception and guilt, and the way these can be recognised as physiological states that can be measured as facial micro-expressions. We have also questioned the design of the experiment that underpins its scientific validation and noted the insufficiency of the sample size of the training data-set used, arguing that the public documentation available provides no indication that the model achieves the performance the developers indicate.

Moreover, even if the model did work as reported, we have highlighted how mass use of deception detection, to be piloted on general populations, is statistically different than targeted use, pointing to the relevance of the rate of false positives in this context. Our analysis demonstrates that the foundational premise of mass screening lacks statistical soundness. This undermines the workings of the iBorderCtrl design in practice. The high-level technopolitical task of detecting irregular border crossings is at best a weak surrogate of the actual machine learning task, that, even in the case of supporting the chain of assumptions in the design, is to reduce a set of facial expressions to a label of deceptive or truthful behaviour when answering a question. This becomes only more startling when we begin to consider the level of complexity of the situations and backgrounds of populations at border-crossings.

By interrogating actual systems in this way, we can begin to question the very premise of their development, moving away from the notion that these systems are fulfilling the tasks they claim to be attending to, and instead ask what function such projects carry out in the creation of subjects and management of populations. This function is not merely technical, but rather, as we have argued, distinctly political, and forms part of a mode of governance increasingly shaping life opportunities and fundamental rights.

## Acknowledgments

Research for this article is part of a large multi-year project called 'Data Justice: Understanding datafication in relation to social justice' (DATAJUSTICE) funded by an ERC Starting Grant (no. 759903). We are grateful to Vera Wilde, Andrew McStay and Nello Cristianini for their comments on earlier drafts of this article.